\documentclass[acmsmall]{acmart}                                    

\usepackage{tabu} 
\usepackage{xcolor}
\usepackage{caption}
\usepackage{subcaption}
\usepackage{import}
\usepackage{todonotes}
\usepackage{multirow}
\usepackage{graphicx}
\usepackage{cleveref}
\usepackage{soul}
\usepackage[normalem]{ulem}

\def \red #1{{\textcolor{black}{#1}}}

\newcommand{\revision}[1]{\textcolor{black}{#1}}
\newcommand{\mnrevision}[1]{\textcolor{black}{#1}}

\AtBeginDocument{%
  \providecommand\BibTeX{{%
    \normalfont B\kern-0.5em{\scshape i\kern-0.25em b}\kern-0.8em\TeX}}}

\setcopyright{acmcopyright}
\copyrightyear{2023}
\acmYear{2023}
\acmDOI{XXXXXXX.XXXXXXX}

\acmJournal{PACMHCI}
\acmVolume{7}
\acmNumber{CSCW1}
\acmArticle{0}
\acmMonth{4}

\begin{document}

\title%
{``I Want to Figure Things Out'': Supporting Exploration in Navigation for People with Visual Impairments}

\author{Gaurav Jain}
\affiliation{
    \institution{Columbia University}
    \country{USA}
}

\author{Yuanyang Teng}
\affiliation{
    \institution{Columbia University}
    \country{USA}
}

\author{Dong Heon Cho}
\affiliation{
    \institution{Columbia University}
    \country{USA}
}

\author{Yunhao Xing}
\affiliation{
    \institution{Columbia University}
    \country{USA}
}

\author{Maryam Aziz}
\authornote{This work was done while Maryam Aziz was an intern at Columbia University.}
\affiliation{
    \institution{University of Connecticut}
    \country{USA}
}

\author{Brian A. Smith}
\affiliation{
    \institution{Columbia University}
    \country{USA}
}


\begin{abstract}

Navigation assistance systems (NASs) aim to help visually impaired people (VIPs) navigate unfamiliar environments. Most of today’s NASs support VIPs via turn-by-turn navigation, but a growing body of work highlights the importance of exploration as well. It is unclear, however, how NASs should be designed to help VIPs explore unfamiliar environments. In this paper, we perform a qualitative study to understand VIPs' information needs and challenges with respect to exploring unfamiliar environments, with the aim of informing the design of NASs that support exploration. Our findings reveal the types of spatial information that VIPs need as well as factors that affect VIPs' information preferences. We also discover specific challenges that VIPs face that future NASs can address such as orientation and mobility education and collaborating effectively with others. We present design implications for NASs that support exploration, and we identify specific research opportunities and discuss open socio-technical challenges for making such NASs possible. We conclude by reflecting on our study procedure to inform future approaches in research on ethical considerations that may adopted while interacting with the broader VIP community.


\end{abstract}




\begin{teaserfigure}
    \centering
      \includegraphics[width=0.95\textwidth]{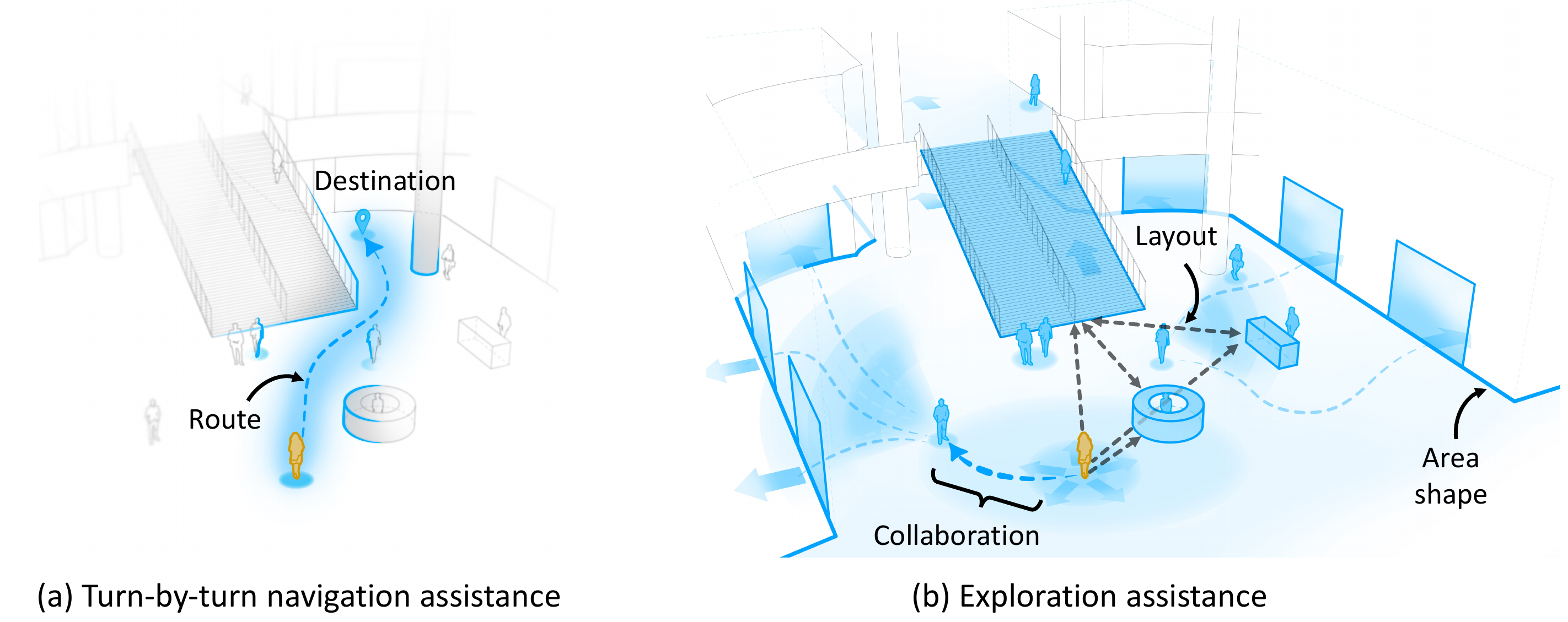}
      \caption{%
      Towards exploration in navigation assistance systems (NASs). (a) Most existing NASs support \textit{turn-by-turn navigation} and give visually impaired people (VIPs) a route-focused view of the world. (b) Our findings reveal how NASs can evolve to be instruments for  \textit{exploration} as well, allowing VIPs to form more complete cognitive maps. Specifically, NASs can facilitate exploration \revision{in both indoor and outdoor environments:} by conveying \revision{area shapes}, by conveying the layout of objects within \revision{environments}, and by facilitating effective collaboration between users and others who might enable even further exploration.
      }
      \label{fig:teaser}
\end{teaserfigure}

\maketitle

\section{Introduction}

\label{sec:intrduction}
Navigating unfamiliar environments is difficult for visually impaired people (VIPs).
\revision{Navigation assistance systems (NASs) are systems that provide users with real-time feedback about their surroundings and their location within the space to enable interactions for supporting navigation.}
Most NASs such as NavCog~\cite{ahmetovic_navcog_2016} and CaBot~\cite{guerreiro_cabot_2019} help VIPs navigate unfamiliar environments via turn-by-turn navigation: by guiding users with verbal instructions (e.g., ``turn left in 2 meters'')~\cite{ahmetovic_navcog_2016, fallah_user_2012, bai_virtual-blind-road_2018,murata_smartphone-based_2018, ran_drishti_2004, riehle_indoor_2008} or by acting as a metaphorical ``guide dog'' that users follow~\cite{guerreiro_cabot_2019, kayukawa_bbeep_2019, kayukawa_guiding_2020, avila_soto_dronenavigator_2017, peng_indoor_2017, saegusa_development_2011, scheggi_cooperative_2014}. 
While turn-by-turn navigation is successful at getting users to their destinations, several researchers have argued that it endorses a passive role for users during navigation --- by having them simply follow directions from the system --- instead of giving users the freedom to explore and understand spaces themselves~\cite{clemenson_rethinking_2021, leshed_car_2008, ingold_being_2011, dey_getting_2018, dey_understanding_2019, munzer_navigation_2012}.


Research in cognitive science has found that the ability to explore and to play an active role in navigation is crucial for one's spatial learning~\cite{chrastil_active_2012, chrastil_active_2014}, cognitive development and brain health~\cite{maguire_navigationrelated_2000, clemenson_improving_2019, konishi_spatial_2013}, sense of independence~\cite{giudice_navigating_2018, golledge_geography_1993, clark-carter_efficiency_1986}, and agency in making wayfinding decisions~\cite{chrastil_there_2015, chrastil_active_2013}. In fact, Banovic et al.~\cite{banovic_uncovering_2013} found that VIPs strongly desire the ability to explore unfamiliar environments, with one participant adding that ``it would be great, magical [to be able to explore] with no one there.'' Yet Banovic et al.\ also found that VIPs hesitate to explore unfamiliar environments without proper assistance due to the challenges that they face when trying to explore them. 
Together, this research shows two opportunities for future NASs to benefit VIPs: by supporting exploration and spontaneity in navigation, and by helping VIPs overcome challenges in exploring unfamiliar environments so they become confident in doing so.

In this paper, we aim to establish a holistic understanding of how NASs can be designed to support exploration. Specifically, we seek to discover the information that VIPs need in order to explore unfamiliar environments, as well as understand the challenges that VIPs face when exploring unfamiliar environments. Hence, we investigate the following two research questions:\\
\vspace{0.05cm}
\begin{tabular}{{p{0.05\textwidth}p{0.85\textwidth}}}
 \textbf{RQ1.} &  What information do VIPs need to explore unfamiliar environments, and what factors influence these needs between individuals? \\
 \textbf{RQ2.} & What challenges do VIPs face while exploring unfamiliar environments, both independently and collaboratively? 
\end{tabular}




Regarding RQ1, past research has explored how turn-by-turn navigation systems might facilitate exploration by providing additional information about their surroundings~\cite{paneels_listen_2013, blum_spatialized_2013, sato_navcog3_2019}
and by providing guidance information in a format that enables active engagement~\cite{ clemenson_rethinking_2021, MSsoundscape}.
Both Banovic et al.'s findings~\cite{banovic_uncovering_2013} and findings from cognitive science~\cite{hersh_mental_2020, long_establishing_nodate}, however, suggest that facilitating exploration requires more than these efforts. 
Specifically, VIPs desire high-level overviews about their surrounding spaces~\cite{banovic_uncovering_2013} and can benefit from the ability to form allocentric cognitive maps of spaces (i.e., ones with environment-centered reference frames) rather than egocentric ones~\cite{long_establishing_nodate}. Through RQ1, we seek to discover what the specific form of that high-level information should be and how that might differ depending on individual VIPs' preferences and backgrounds.

Regarding RQ2, 
a growing body of work prescribes that NASs should not be viewed as merely granting VIPs independence but rather should be viewed as one partner in a collective interdependence effort~\cite{bennett_interdependence_2018}, improving the collaboration between VIPs and others around them. 
As a result, it is important to understand exploration challenges that arise not only from VIPs' own abilities, but also from failures in how others collaborate with VIPs.

To investigate RQ1 and RQ2, we perform 
a series of in-depth semi-structured interviews with 12 VIPs and short follow-up interviews with six other people with whom the VIPs often collaborate \revision{by employing
a recent Critical Incident Technique (CIT)~\cite{flanagan_critical_1954}}. These six people included orientation and mobility (O\&M) trainers, store employees, and leaders of blind-serving organizations.
In addition, \revision{our research team included two VIP interns} who helped us connect with participants, develop our interview protocol, and interpret the study data.

Through our interviews, we discovered that VIPs need two types of spatial information for exploration --- \textit{shape} information and \textit{layout} information --- in addition to the information they need for turn-by-turn navigation (i.e., POIs and route information). 
We also identified three factors that influence individuals' preferences for spatial information: 
(1) their onset of vision impairments, (2) their inherent sociability, and (3) their O\&M proficiency and mobility aid preference. Understanding \textit{how} these factors dictate the format, source, and amount of spatial information VIPs need leads to an inclusive design of future NASs that support VIPs' exploration.


Regarding VIPs' challenges in exploration, we found that VIPs hesitate to explore environments independently because of their lack of confidence in acting upon spatial information collected via non-visual senses (e.g., hearing, touch, smell). Moreover, even when reliable spatial information is available, VIPs still do not feel they can explore environments independently because their self-reported O\&M proficiency falls short. When collaboratively exploring environments, both VIPs and their collaborators reported that social pressure on the person helping the VIP is a major obstacle to their successful collaboration. In addition, we found that VIPs' find it difficult to supervise and comprehend assistance from their collaborators. 



Figure~\ref{fig:teaser} summarizes how our findings can inform the design of future NASs. Figure~\ref{fig:teaser}a represents the state of NASs today: they are oriented around turn-by-turn navigation and give VIPs a route-focused view of the world, surfacing only those details about VIPs’ greater surroundings that are adjacent to VIPs’ routes.
Figure~\ref{fig:teaser}b, on the other hand, represents how NASs can evolve to offer exploration assistance as well, leading to a greater formation of cognitive maps. They can facilitate exploration in several ways: by conveying \revision{area} shapes, by conveying the layout of objects within \revision{environments}, and by facilitating effective collaboration between users and others who might enable even further exploration. We describe these possibilities in detail in our Discussion section where we identify specific research opportunities\revision{, discuss socio-technical challenges, and analyze practical considerations}; that can bring us closer to this broader vision for NASs. \revision{Finally, we inform future approaches in research within the CSCW community on ethical considerations that may be adopted while engaging with the broader VIP community.}

\section{Related Work}

Research in cognitive science (Section~\ref{sec:cog-sci-RW}) has shown the importance of exploration to navigation~\cite{chrastil_active_2012, maguire_navigationrelated_2000, clemenson_improving_2019, konishi_spatial_2013, giudice_navigating_2018, golledge_geography_1993, clark-carter_efficiency_1986, chrastil_there_2015, chrastil_active_2013, tolman_cognitive_1948, okeefe_hippocampus_1978, golledge_wayfinding_1999, ggolledge_cognitive_2000}, 
yet most NASs focus on guiding VIPs via turn-by-turn navigation~\cite{ahmetovic_navcog_2016, fallah_user_2012, bai_virtual-blind-road_2018,murata_smartphone-based_2018, ran_drishti_2004, riehle_indoor_2008, guerreiro_cabot_2019, kayukawa_bbeep_2019, kayukawa_guiding_2020, avila_soto_dronenavigator_2017, peng_indoor_2017, saegusa_development_2011, scheggi_cooperative_2014}. 
Although recent NAS approaches have taken important first steps toward facilitating exploration for VIPs~\cite{blum_spatialized_2013, paneels_listen_2013, MSsoundscape, sato_navcog3_2019}, many questions remain.
Here, we summarize the evolution of exploration in NASs (Section~\ref{sec:NAS-design-approaches}) and work on understanding VIPs' navigational needs and challenges
(Section~\ref{sec:understanding-needs-prefs}), describing how the present work advances these efforts.

\subsection{The Importance of Exploration in Navigation}
\label{sec:cog-sci-RW}
\revision{
Exploration involves incidental learning of a space which supports the ability to make in-situ navigation decisions~\cite{ggolledge_cognitive_2000}. While one gathers both \textit{route} knowledge and \textit{survey} knowledge when navigating environments, exploration is primarily supported via survey knowledge~\cite{golledge_wayfinding_1999}. Route knowledge includes path from one point to another and landmarks along the path and is acquired via an egocentric reference frame (person-centered). In contrast, survey knowledge includes configuration of the space and is acquired via an allocentric reference frame (environment-centered).
}

Various strands of cognitive science research indicate the importance of exploration to navigation. Specifically, the ability to explore environments leads to two primary cognitive benefits: the formation of cognitive maps~\cite{okeefe_hippocampus_1978, tolman_cognitive_1948} and the promotion of active engagement with the environment~\cite{chrastil_there_2015, chrastil_active_2013}. Regarding the formation of cognitive maps, Tolman et al.~\cite{tolman_cognitive_1948} 
found that maps created via exploration 
can be accessed and used from any location in the environment and not just from specific places or routes. 
Regarding the promotion of active engagement, Chrastil et al.~\cite{chrastil_there_2015} report that playing an active role in spatial decision-making leads to a better spatial understanding of the environment. This understanding can be useful in many situations, such as when the need to find an alternative route arises~\cite{chrastil_active_2013}.
In addition to these two benefits, many researchers have shown that the ability to explore unfamiliar environments can improve the development of cognitive abilities and improve overall brain health~\cite{maguire_navigationrelated_2000, clemenson_improving_2019, konishi_spatial_2013}.

This body of work reveals
the potential benefits that NASs could bring by facilitating exploration for VIPs. Since we do not yet understand how exactly NASs should be designed to support exploration for VIPs, our work aims to address this research gap.


\subsection{\revision{Tools for Supporting Exploration in Navigation}}
\label{sec:NAS-design-approaches}

\revision{
Various navigation tools support VIPs' exploration in unfamiliar environments.
Tactile maps have been shown to provide VIPs with a spatial representation of an environment via touch~\cite{espinosa_using_1998, trinh_feeling_2020, abd_hamid_facilitating_2013, jacobson_navigating_1998, taylor_customizable_2016, nadel_hippocampus_1980, pissaloux_accessible_2018}, enabling them to preview an unfamiliar environment before their visit~\cite{blades_map_1999, miesenberger_pre-journey_2014}.
Another set of tools use virtual reality (VR) for developing virtual environment systems that support VIPs' learning and exploration of both virtual~\cite{siu_virtual_2020, zhao_seeingvr_2019} and physical environments~\cite{lahav_blindaid_2008, todd_virtunav_2014, zhao_enabling_2018}.
Many tools such as AIRA~\cite{aira}, VizWiz~\cite{bigham_vizwiz_2010}, and BeMyEyes~\cite{bill_holton_review_2015} utilize crowdsourcing to connect VIPs with sighted people over the internet for real-time remote sighted assistance. 
}


\revision{%
Within tools that support exploration, our work focuses on navigation assistance systems (NASs). Recall that we defined NASs as systems that provide users with real-time feedback about their surroundings and their location within the space to enable interactions for supporting navigation. Most NASs follow a turn-by-turn navigation approach to guide users to their destination via verbal descriptions~\cite{ahmetovic_navcog_2016, fallah_user_2012, bai_virtual-blind-road_2018,murata_smartphone-based_2018, ran_drishti_2004, riehle_indoor_2008} or by acting as a metaphorical ``guide dog'' that users follow~\cite{guerreiro_cabot_2019, kayukawa_bbeep_2019, kayukawa_guiding_2020, avila_soto_dronenavigator_2017, peng_indoor_2017, saegusa_development_2011, scheggi_cooperative_2014}.
}%

Several recent works in human--computer interaction (HCI) have argued for NASs to facilitate exploration in addition to turn-by-turn guidance. 
Nair et al.~\cite{Nair2021a} recently highlighted the importance of exploration and the ability to look around when navigating, creating a system called NavStick for giving VIP players the ability to look around within adventure video games. 
\revision{Dey et al.~\cite{dey_getting_2018, dey_understanding_2019} show that even simple modifications in existing NAS-interfaces that go beyond just providing turn-by-turn directions can help sighted people acquire spatial knowledge of an environment.
}
Clemenson et al.~\cite{clemenson_rethinking_2021} recently argued that NAS designers should rethink the role of turn-by-turn navigation systems so that they can facilitate active engagement with the world.

NAS developers have also made efforts to facilitate exploration~\cite{blum_spatialized_2013, paneels_listen_2013, sato_navcog3_2019, MSsoundscape, gleason_footnotes_2018}. 
In-Situ Audio Services (ISAS)~\cite{blum_spatialized_2013} and NavCog3~\cite{sato_navcog3_2019}, for example, announce nearby points of interests (POIs) via speech and 3D-spatialized audio either during a route (via ``pushes'') or via search (``pulls''). Microsoft Soundscape~\cite{MSsoundscape} places audio beacons directly at the destination instead of giving them instructions at every turn, allowing users to determine their own routes. 

Although recent research has emphasized the crucial role that exploration plays in navigation and has made some notable contributions towards it, research in accessibility and cognitive science shows there is more to exploration than learning nearby POIs and the routes to destinations. Exploration in navigation includes learning about spaces and building an environment-centered cognitive map of spaces to make navigation decisions from~\cite{banovic_uncovering_2013, hersh_mental_2020, long_establishing_nodate, siegel_development_1975}.
Before our community can ever develop NASs to support exploration, it is important for us to take a step back and gain a holistic understanding about what information VIPs need in order to explore and why VIPs find it difficult to gather this information in the first place.
Hence, our work does not introduce new technology, but instead aims to broaden our understanding about VIPs' needs and challenges around exploration.

\subsection{Understanding VIPs' Navigation Needs and Challenges}
\label{sec:understanding-needs-prefs}

Several studies have identified VIPs' information needs while navigating \cite{ahmetovic_impact_2019, williams_pray_2013, quinones_supporting_2011, lewis_hearing_2015, van_der_bie_guiding_2016, scheuerman_learning_2017, guerreiro_virtual_2020} and the challenges that VIPs face~\cite{giudice_use_2020, williams_just_2014, easley_lets_2016, thieme_i_2018, langdon_better_2016}, but most of them are aimed at informing the design of NASs that support turn-by-turn navigation. Our focus, on the other hand, is to understand VIPs' information needs and challenges regarding \textit{exploring} unfamiliar environments, which remains understudied~\cite{banovic_uncovering_2013}.

Lewis et al.~\cite{lewis_hearing_2015} interviewed a focus group of VIPs to determine user requirements for NASs, but only in the context of turn-by-turn navigation.
\revision{Specifically, most of their findings correlate with information about POIs, updates on progress about routes to the destination, and preference for notifications about whether users are following the system-generated routes or not. 
Unlike our work, which focuses on NASs that support exploration, Lewis et al.~\cite{lewis_hearing_2015} made proposals to \textit{guide} users to a destination instead of giving them the freedom to choose their own routes.}
Similarly, Scheuerman et al.~\cite{scheuerman_learning_2017} investigate how VIPs write navigation instructions to each other to better inform the design of turn-by-turn navigation systems. \revision{The authors study how VIPs represent distances, indicate directions, and reference landmarks; with an aim to propose design implications for NASs. 
}

Banovic et al.~\cite{banovic_uncovering_2013} interview both VIPs and O\&M trainers to uncover VIPs' information needs for spatial learning. 
Our work builds on Banovic et al.'s work in two ways. First, we shift our focus from spatial learning to in-situ exploration. In other words, we envision NASs helping users to ``figure things out'' on their own and make navigation decisions on the fly instead of learning a space to visit later on. 
Second, we investigate the exploration affordances that NASs should offer on a more specific level (RQ1). Banovic et al.\ found that VIPs desire a ``high-level overview'' of spaces, while we find in Sections~\ref{sec:RQ1-findings} and \ref{sec:RQ2-findings} that NASs should afford \textit{\revision{area} shape} information, \textit{layout} information, and assistance in communicating needs with others specifically, as Figure~\ref{fig:teaser}b illustrates.

Researchers have found that
VIPs' navigation strategies are different from sighted people's~\cite{swobodzinski_indoor_2009} and that
variations in individual VIPs' navigation preferences and behaviors play an important role in designing NASs~\cite{ahmetovic_impact_2019, williams_pray_2013}. 
\revision{For instance, Williams et al.~\cite{williams_pray_2013} identify personality and scenario attributes that impact VIPs' daily-life navigation preferences. Ahmetovic et al.~\cite{ahmetovic_impact_2019} made proposals for NASs to adapt to the verbosity of navigation instructions to suit user's level of expertise, ensuring that they stay engaged in the during navigation.}
As a result, as part of our RQ1 investigation,
we also uncover the factors that affect VIPs' choice of spatial information (Section~\ref{sec:factors-info-needs}) \revision{to identify factors that influence VIPs preferences for spatial information and to understand which specific dimensions of spatial information does each of these factors dictate.}

Last, prior work has identified many \textit{social} challenges related to navigation that VIPs face as well: specifically, challenges in comprehending assistance from other people~\cite{williams_just_2014, guerreiro_airport_2019}, in navigating through environments that are predominantly occupied by sighted people~\cite{easley_lets_2016}, and in negotiating these differences with other people~\cite{thieme_i_2018}. 
VIPs' ability to explore is best viewed within an interdependence framework~\cite{bennett_interdependence_2018, vincenzi_interdependence_2021} --- that is, as a collective effort involving other people, as Figure~\ref{fig:teaser}(b) illustrates.
In this spirit, in RQ2 we investigate not only VIPs' internal challenges when trying to explore, but also how VIPs interact with people when trying to explore. 
Our aim with this latter aspect of RQ2 is to understand how NASs can bridge communication gaps and foster collaboration as yet another means of exploration.

\section{Methods}
We performed a qualitative study guided by our two research questions RQ1 and RQ2.
Here we describe our recruitment process (Section~\ref{sec:ptcpts}), experimental procedure (Section~\ref{sec:procedure}), and data analysis procedure (Section~\ref{sec:dataAnalysis}).

\subsection{Participants}
\label{sec:ptcpts}
We recruited 12 visually impaired participants (VIPs) --- seven men and five women --- with a wide range of ages (18 to 72). Table~\ref{tab:participants} summarizes participants' demographics; all participants' names in the table are pseudonyms. \revision{
Note that we assigned pseudonyms by randomly sampling common names from our participants' geographical locations and racial identities.
}
\revision{Participants identified themselves with diverse racial identities --- specifically, as Asians, Black or African Americans, Whites, or Hispanic/Latinos.} 
\revision{We collected gender information as a free-response, optional field to ensure inclusiveness of all gender identities.}
Participants also had diverse visual abilities, onset of vision impairment, and mobility aid preferences. 
Ten are white cane users and two use guide dogs.
All live in the United States, spread across nine different states.

We recruited the VIPs from several sources. 
First, we posted to online forums and social media platforms popular among the visually impaired community, recruiting six participants in the process. Second, we connected with two participants through our research team's ongoing collaboration with a major blind-serving organization. We recruited the last four participants via snowball sampling~\cite{goodman_snowball_1961}.

\begin{table*}[t]
\caption{
    \textcolor{black}{Self-reported demographic information of our visually impaired participants.}
    }
    \label{tab:participants}
\resizebox{\linewidth}{!}{
\begin{tabular}{lcclll}
\hline 
\textbf{Pseudonym} & \textbf{Gender} & \textbf{Age} & \textbf{Onset} & \textbf{Mobility Aid} & \textbf{Nature of Vision Impairment}                                           
\\ \hline
\revision{Harper}  & F & 59 & Age 20    & Guide dog  & Total blindness \\
\revision{Yusuf}       & M               & 22  & Birth & White cane   & No vision in left eye; Shape, color, and movement in right \\
\revision{Anika} & F & 72 & Age 26 & Guide dog & Total blindness \\
\revision{Booker}   & M & 25 & Age 4     & White cane & Total blindness \\
\revision{Daniel}  & M & 21 & Birth & White cane & Total blindness \\
\revision{Rahul}     & M & 33 & Birth & White cane & Total blindness \\
\revision{Anya}       & F               & 19  & Birth & White cane   & Light perception, shapes, color in left eye; No vision in right       \\
\revision{Miguel} & M & 23 & Birth & White cane & Total blindness \\
\revision{Charles}       & M               & 25  & Age 1     & White cane   & Light, shape, color, and movement in left eye; No vision in right \\
\revision{Zendaya}    & F & 18 & Birth & White cane & Total blindness \\
\revision{Lucas}   & M & 36 & Birth & White cane & Total blindness \\
\revision{Olivia}      & F               & 21  & Birth & White cane   & Light perception in left eye; Color, contour, and shadow in right     \\ \hline
\end{tabular}
}
    
\end{table*}

In addition to the VIPs, we recruited six other participants from groups who the VIPs reported as often collaborating with and who play a big part in shaping VIPs' behaviors around navigation. The six others include two leaders of a major blind-serving organization (their assistive technology director and their director of O\&M services), a manager of disability services at a large university, an O\&M trainer, and two store employees (one working at a small bookstore and the other at a major department store).
We selected these two store employees because they self-reported as regularly interacting with visually impaired customers.

\subsection{Procedure}
\label{sec:procedure}
\revision{
To develop our interview protocol and study procedure, we conducted pilot interviews and brainstorming sessions with two VIPs --- working as paid summer research interns as co-authors on another project within our lab --- in order to begin understanding VIPs' experiences around exploration. Both VIP interns reported being partially blind (or low vision), and were college students (both males, aged 20 and 21). We invited them to serve as pilot study participants, to review our interview questions, and to discuss participants' responses. Involvement of the two VIP interns helped us in many ways. First, during our pilot interviews with them, they provided us an initial understanding about VIPs' experiences around exploration, helping us refine the interview questions and removing ambiguous language from it. Second, they helped us recruit participants via snowball sampling~\cite{goodman_snowball_1961}. Last, being able to consult them for quick clarifications about our understanding of the study data was crucial to the data analysis phase.
We reflect on the ethical considerations of involving the two VIP interns in our study in Section~\ref{sec:ethical_cons}.
}%

Our final interview protocol was organized as follows.
We began each interview by collecting participants' background information, including their demographics (age, gender, race, location), onset and level of vision impairment, history of O\&M practices, mobility aid preference, and use of assistive technologies for navigation.
Collecting this background information helped us identify correlations between different individuals' exploration preferences and their past experiences.
We then turned to RQ1 and RQ2 by employing
a recent Critical Incident Technique (CIT)~\cite{flanagan_critical_1954}, in which we asked participants to recall \revision{and describe a recent time (or \textit{incident}) where they explored an unfamiliar environment.
CIT is commonly employed by the HCI and CSCW community for conducting in-depth interviews~\cite{hsueh_deconstructing_2019, kelly_demanding_2017, nouwens_application_2018, wong_collaboration_2017}. 
We adopted CIT because it gave us actual examples of times when our participants faced challenges in exploring unfamiliar environments, without having to undergo an extensive contextual inquiry and follow participants around all day long. 
However, as we discuss later in limitations (Section~\ref{sec:limitations}), future research could shadow VIPs to understand the context of their experiences in more detail and reveal more subtle nuances that may have been overlooked by using CIT.}

\revision{%
Between our participants, we learned about many incidents of navigating unfamiliar \textit{indoor} environments (e.g., hotels, grocery stores, airports, train stations, academic buildings at universities) and \textit{outdoor} environments (e.g., sidewalks, street crossings, large open spaces in shopping complexes). Examples of incidents that participants shared include traveling from their hotel rooms to a vending machine or to breakfast, walking around a grocery store to find a specific item which was moved to a different location by management, making their way from a parking lot to a specific department at a university, crossing a busy street by themselves, exploring a shopping complex, and figuring out their way to a recently opened café. Importantly, we noticed that participants tended to explore unfamiliar outdoor environments less frequently than indoor spaces due to them being more concerned about their safety in outdoor settings.
}



\revision{
For each recent incident that the participants mentioned, we asked follow-up questions to help them recall how they transpired and to help our team better understand their behaviors during the incident. In general, we led the conversation with questions such as: \textit{"What did you look for when navigating in the space?"}, \textit{"What strategies did you use to gather this information?"}, \textit{"What were the challenges that you faced while doing so?"}, \textit{"What other pieces of information would have helped navigate this environment?"}, \textit{"Were you navigating by yourself?"}, \textit{"Did you consider reaching out for help?"}, \textit{"Who helped you, and how?"}, and \textit{"Which aspect of gathering information did they assist you with?"}}
Through these discussions, we learned about participants’ information needs (RQ1), challenges (RQ2), and workarounds to address their needs and challenges.




During the interviews, we also made special note of people who the VIPs mentioned as often collaborating with and their effect on the VIPs' navigation behaviour. To further investigate RQ2 from the perspective of these other people, we conducted short follow-up interviews with six representative non-VIP participants, described in Section~\ref{sec:ptcpts}.
Soliciting experiences from both VIP and non-VIP perspectives helped us find the root cause of the challenges that VIPs --- and in some case these non-VIPs --- face when exploring unfamiliar environments collaboratively.

We asked these six others about their backgrounds, their frequency of interactions with VIPs, their role in assisting VIPs in navigation, and their experience with assistive technologies for VIPs (if any). We used the same critical incident technique~\cite{flanagan_critical_1954} as before, but we asked these participants to recall and walk us through their past experiences interacting with VIPs, specifically instances in which they helped VIPs with navigation-related tasks.

We conducted all interviews via a Zoom video-conference, except for two: the bookstore employee preferred to be interviewed in person during working hours, and Anika preferred a telephone interview since she was not comfortable using Zoom. All interviews were recorded with participants' consent.
Each interview lasted 60--90 minutes and the participants were compensated \$25 for their time. We collected over \textcolor{black}{23 hours} of recordings in total. The study was approved by our institution's IRB. 

\subsection{Data Analysis}
\label{sec:dataAnalysis}
We analyzed our interview data via a grounded theory approach~\cite{charmaz_constructing_2006}. Following grounded theory, we simultaneously performed data analysis and collection, iteratively refining our analytic frame while also updating the questions for future interviews as we identified emerging themes.
The primary author and a co-author collaboratively performed open coding~\cite{strauss_basics_1990} on the interview transcripts while interviews were still ongoing to arrive at an initial set of codes. 
We consulted the two VIP interns to validate and refine some of these themes. \revision{We synthesized the initial set of codes and recruited participants for interviews as necessary to arrive at theoretical saturation~\cite{strauss_basics_1990}, which we achieved after interviewing 12 participants.}
Finally, we conducted weekly meetings to review the codes and refine them into a closed set of codes, which we then re-applied to the transcripts to reach a ``group consensus''~\cite{saldana_coding_2013}.
\label{sec:findings}

\section{RQ1 Findings: Information Needs for Exploration}
\label{sec:RQ1-findings}
In this section, we present our findings for RQ1: What information do VIPs need to explore unfamiliar environments, and what factors influence these needs between individuals? 
Our interviews revealed that VIPs need two types of spatial information (Section~\ref{sec:informationNeeds}) for exploration --- shape information and layout information --- in addition to the information they need for turn-by-turn navigation (i.e., POIs and route information). 
We also discovered factors that influence these information needs between individuals 
(Section~\ref{sec:factors-info-needs}).
Throughout our results, we highlight key insights that inform the design of future NASs supporting VIPs' exploration in navigation.



\subsection{Spatial Information}
\label{sec:informationNeeds}

We found that participants want to learn spatial information in a hierarchy --- first, the shape information, then the layout information, followed by specific details about the route to a destination. Participants noted that the shape and layout information give them a high-level overview of the environment, allowing them to decide their route to the destination by themselves. \red{Our findings extend prior work~\cite{banovic_uncovering_2013} by specifying what it means for VIPs to obtain a ``high-level'' overview of a space.}

We note an important difference between what VIPs want during exploration and how current NASs facilitate exploration. \revision{While current NASs' approach of announcing nearby POIs~\cite{paneels_listen_2013, blum_spatialized_2013, sato_navcog3_2019} promotes a egocentric reference frame, we found that VIPs want to build their cognitive maps in an allocentric reference frame via shape and layout information.}
\revision{Next, we describe the two types of spatial information.}

\subsubsection{Shape information}
\label{sec:shapeInfo}
Participants described shape information as a skeletal wire-frame of the unfamiliar environment, marking its bounds. Shape information provides VIPs with a high-level overview of the structure of the space very quickly.
Many participants (n=4) indicated that shape information forms the basis of their mental map, onto which they append other pieces of information including object layouts and routes. \revision{Booker}, recollecting his recent visit to an office complex, explained how he would ask for shape information before anything else:
\vspace{-0.25cm} \\\\
\begin{tabu} to \linewidth { X[l] X[28l] X[l]}
 & \textit{``[I ask] ... `What does this building generally look like?' Because I want to understand its shape. That’s one of the first things [I want] to understand.''} --\revision{Booker}  & \\
\end{tabu} \vspace{-0.1cm} \\

\red{Participants described using shape in context of both indoor and outdoor environments. For indoors, shape is understood to be the boundary or walls of a room and thus, can be represented in the form of geometric shapes such as squares (e.g., offices) or rectangles (e.g., hallways).
For outdoors, participants referred to the environment's general dimensions of walkable area as its shape. For instance, \revision{Olivia} explained how she values information about width of the sidewalk and the streets when the environment may require crossing streets.}

%
%

Interestingly, we learned that VIPs use shape information differently for turn-by-turn navigation and exploration. \red{Williams et al.~\cite{williams_just_2014} found that, for turn-by-turn navigation, VIPs use shape information to orient themselves and walk in straight lines. In contrast, we found that, for exploration, VIPs use shape information to get a sense of the environment's high-level structure.}

\subsubsection{Layout information}
\label{sec:layoutInfo}
Participants described layout information as the arrangement of different objects within the space --- both with respect to them (egocentric)  and with respect to the environment (allocentric). The objects constitute several categories including points of interests (POIs) (e.g., stairs, elevators, restrooms), landmarks (e.g., change in flooring, curbs), obstacles, and other people in the environment who the VIPs often see as potential collaborators in exploration. 

For example, \revision{Yusuf} recalled his visit to an unfamiliar grocery store where he first focused on understanding the shape of the store, and then populated this structure with layout information.
\revision{He explained that he sought to understand the locations of certain POIs such as the bathroom, specific aisles, and the cash register to develop a more comprehensive understanding of the space, enabling him to explore the grocery store}. \revision{Charles} described the layout of a space he recently visited with the help of verbal descriptions from his friend: \vspace{-0.25cm} \\\\
\begin{tabu} to \linewidth { X[l] X[28l] X[l]}
 & \textit{``... it's two floors, there's a restroom down the hall on the way as soon as you get in, there is a large step up on to the porch. And then there is double doors leading out to the pool. And there's no gate around the pool.''} --\revision{Charles}  & \\
\end{tabu} \\

We noticed subtle differences in VIPs' use of landmarks and POIs between navigating to a destination and exploring unfamiliar environments. On one hand, participants described using landmarks and POIs either as potential destinations or as locations to reorient themselves along the route; pointed out by prior work ~\cite{giudice_navigating_2018}. For exploration, on the other hand, participants mentioned using landmarks to create a more comprehensive mental map \revision{so that they can} make navigation decisions in situ.

\vspace{2mm}
\begin{center}
\setlength{\fboxsep}{0.8em}
\fbox{\begin{minipage}{0.5\textwidth}
\textbf{Insight 1}: VIPs need two types of spatial information: shape information and layout information, to get a high-level overview of a space.
\end{minipage}}
\end{center}
\vspace{1mm}


\subsection{Factors Influencing Information Preferences Between Individuals}
\label{sec:factors-info-needs}
We discovered three factors that influence VIPs' preferences for spatial information: 
 (1) their onset of vision impairments, (2) their inherent sociability, and 
(3) their O\&M proficiency and mobility aid preference. \red{Not surprisingly, many of these factors have been observed to affect VIPs' navigation behavior in different contexts ~\cite{long_establishing_nodate, wiener_foundations_2010, schinazi_spatial_2016, thinus-blanc_representation_1997, williams_just_2014, williams_pray_2013}. Thus, we identified dimensions of information needs --- specifically, spatial information --- dictated by these factors in context of exploring unfamiliar environments.}
VIPs' preferences dictated information needs along three dimensions, (1) preferred format of spatial information: sensory augmentation vs. detailed descriptions, (2) preferred source of spatial information: via other people vs. via self, and (3) amount of support for gathering spatial information: minimal vs. maximal support; respectively.





\subsubsection{Onset of vision impairment and preferred `format' of spatial information}
\label{sec:spatial-info-format}

We observed differences in VIPs' preferred \textit{format} for spatial information, which depended on when their onset of vision impairment came about.

Many early-blind participants (n=3) --- people who are blind by birth or developed vision impairments early in life --- shared that they have learned to trust their non-visual senses to collect spatial information over time. These participants further explained that they prefer receiving \textit{confirmation} of their sensed information either from other people or through their other senses, instead of receiving complete \textit{descriptions}, and that they prefer their non-visual senses to be amplified (e.g., through a more powerful echolocation ability) rather than mutated (e.g., through system-provided labels). \red{These participants noted that doing so helps them maintain their sense of independence, which is important to VIPs as reported by Lee et al.~\cite{lee_designingIndep_2021}.}

By contrast, the late-blind participants expressed preference for receiving spatial information in the form of verbal descriptions with as much visual detail as possible.
Two of our late-blind participants, \revision{Anika} and \revision{Harper}, attributed their preference for descriptions of spatial information to their having had vision in the past, allowing them to visualize spaces. \vspace{-0.25cm}\\\\
\begin{tabu} to \linewidth { X[l] X[28l] X[l]}
 & %
 \textit{``I have a pretty good running movie of the world and what it looks like around me and a picture of it. [T]his knowledge [of] how things are put together and look like is definitely [an] advantage of having seen for 25 years.''}  --\revision{Anika} 
 & \\
\end{tabu} \\\\
\revision{Harper noted that including visual details, such as colors, within verbal descriptions helped her visualize spaces and memorize them}. None of the early-blind participants showed interest in learning or referring to visual concepts such as color when receiving spatial information.






\subsubsection{Inherent sociability and preferred `source' of spatial information}
We found that participants' preference for seeking assistance from others to collect spatial information --- i.e., \textit{collaborative exploration} --- is correlated to their inherent sociability and prior experiences of being assisted by others. 
While all participants described instances of seeking assistance from others, many (n=5) expressed their preference for using other people as a source of spatial information, instead of doing it all by themselves.

For some of these participants it was easier (\revision{Miguel}) and faster (\revision{Lucas}) to gather information via others. Many others (n=4) felt independence in seeking help from others, stating that \textit{``independence also has to do with self-advocating for [oneself]''} (\revision{Zendaya}). Most participants who preferred collaborative exploration shared positive experiences of receiving help, even from strangers, and self-reported as being socially adept. 

Interestingly, \revision{Anika} mentioned that asking for spatial information from others was also a way for her to connect with people. 
\revision{She shared instances of building relationships with several people who help her on a regular basis:}
\vspace{-0.25cm}\\\\
\begin{tabu} to \linewidth { X[l] X[28l] X[l]}
 & %
 \textit{`` I could do online shopping or delivery shopping with CVS, but I don't want to. These people become my friends. 
 [...] 
 I like going up and connecting with the people.''} --\revision{Anika}
 & \\
\end{tabu} \\


Other participants (n=4) expressed their preference for gathering spatial information by themselves --- i.e., \textit{independent exploration} --- whenever it was possible. For most of these participants, we found that they value their ability to explore independently above everything else, and some expressed discomfort in interacting with other people due to negative experiences of seeking help from others. \red{We heard many stories about people not knowing how to help VIPs, similar to what Williams et al.~\cite{williams_just_2014} and Guerreiro et al.~\cite{guerreiro_airport_2019} have reported.}



\subsubsection{O\&M proficiency/mobility aid preference and preferred `amount of support'}
Participants were split on the amount of support they need to gather spatial information in order to successfully explore unfamiliar environments. These differences were found to be correlated with VIPs' self-reported proficiency in O\&M skills (being exposed to limited/no training vs. sufficient training) and their mobility aid preference (guide dog vs. white cane).

Most participants who self-reported as having limited exposure to O\&M training (n=3) described wanting a higher degree of support from NASs, i.e., detailed spatial information, compared to those who had a greater exposure to O\&M training. In fact, some of the participants with limited O\&M exposure reported to be content with turn-by-turn navigation and did not express a strong desire for the ability to explore unfamiliar environments. As a consequence, they found it extremely hard to infer spatial information on their own.
\vspace{-0.25cm} \\\\
\begin{tabu} to \linewidth { X[l] X[28l] X[l]}
 & %
 \textit{``Usually for groceries, I ask somebody to assist me because [...] I feel like the shape of the store is really complicated to learn.''} --Miguel
 & \\
\end{tabu} \\

We also observed a similar pattern with mobility aid preference. Both of our participants who use a guide dog --- \revision{Anika} and \revision{Harper} --- exclaimed that it is extremely hard for them to explore environments with the help of a guide dog. 
\revision{\revision{Harper} explained that exploring environments with a guide dog is more challenging than with a white cane.}\vspace{-0.25cm} \\\\
\begin{tabu} to \linewidth { X[l] X[28l] X[l]}
 & %
 \textit{\revision{``That's the real downside for my seeing eye dog. To explore a new space, I am supposed to tell him, give him commands. If I don't know where I'm going, I can't tell him.''} --\revision{Harper}} 
 
 & \\
\end{tabu}
\vspace{-0.4cm}\\\\

\vspace{2mm}
\begin{center}
\setlength{\fboxsep}{0.8em}
\fbox{\begin{minipage}{0.5\textwidth}
\textbf{Insight 2}: VIPs' preferred format, source, and amount of spatial information support vary between individuals depending upon their \mbox{onset} of vision impairment, their inherent sociability, and their O\&M proficiency and mobility aid \mbox{preference}, respectively.
\end{minipage}}
\end{center}
\vspace{1mm}


\section{RQ2 Findings: Challenges in Exploration}
\label{sec:RQ2-findings}

In this section, we present our findings for RQ2: What challenges do VIPs face while exploring unfamiliar environments, both independently and collaboratively? 
Our analysis revealed several challenges that VIPs face during exploration, which we present around two major means of exploration, (1) \textit{independent exploration}: exploring by themselves (Section~\ref{sec:ind_exploration}), and (2) \textit{collaborative exploration}: exploring by seeking assistance from others (Section~\ref{sec:colab_exploration}). 

\subsection{Challenges in Independent Exploration}
\label{sec:ind_exploration}
We discovered two major challenges in our participants' independent exploration endeavors.
First, participants reported difficulties in gathering spatial information precisely and reliably via their non-visual senses (e.g., hearing, touch, smell). Second, participants described challenges in acquiring appropriate O\&M training and maintaining their O\&M skills. We elaborate upon these challenges in the following sections.



\subsubsection{Gathering spatial information precisely and reliably}
\label{sec:precise-spatial-info}



Interviewees shared many instances in which they gathered spatial information --- shape and layout information (see Section~\ref{sec:informationNeeds}) --- using their non-visual senses such as hearing, touch, and smell. \red{Not surprisingly, VIPs' use of their non-visual senses for navigation has been reported by prior research as well~\cite{thinus-blanc_representation_1997, banovic_uncovering_2013, williams_pray_2013, williams_just_2014, thieme_i_2018}. Interestingly, though, we found that spatial information gathered through these senses was neither precise nor reliable for VIPs to make in-situ navigation decisions, severely impeding their ability to explore unfamiliar environments independently.}


Some participants (n=3) described using their sense of hearing --- particularly, echolocation --- to gather shape information, but lamented that it only gives them a very rough idea of the environment's shape and lacks the precision to work well in large or crowded places. 
\revision{Booker recalled visiting a shopping mall, describing how echolocation was not precise enough for him to gather shape information.} \vspace{-0.25cm}\\\\
\begin{tabu} to \linewidth { X[l] X[28l] X[l]}
 & %
 \textit{``... echolocation can tell me some things about the shape, but I want specifics. 
   So, I will go to each side of the entrance and check how big the entrance is, and then I can learn that there are two large openings from this foyer.''} --\revision{Booker}
 & \\
\end{tabu}
\\


\revision{Although our participants reported trusting their sense of touch over other senses, it was still not a reliable way as the ``brute force approach'' does not scale well to large spaces (e.g., airports, shopping complexes)}.
Furthermore, participants expressed hesitation in using touch, including tactile feedback via their cane, to gather spatial information when other people are present in the environment. 
\revision{Charles} mentioned concerns about hitting people with his cane in crowded places, while \revision{Olivia} expressed fear of knocking objects over in front of others. \revision{Harper echoed the sentiment, stating that using touch might be inappropriate in some public settings.}
 \vspace{-0.25cm} \\\\
\begin{tabu} to \linewidth { X[l] X[28l] X[l]}
 & %
 \textit{``I mean, if you're somewhere by yourself, you could explore anything. But when you're in a public setting, people are really gonna look at you strangely if you get up at a restaurant and start walking around and feeling all the tables.''} --\revision{Harper}
 & \\
\end{tabu}
\vspace{-0cm}\\\\
Approaching this finding from Easley et al.'s~\cite{easley_lets_2016} perspective --- the majority of people occupying an environment dictate its social norms --- we could in fact say that VIPs' independent exploration strategies using their non-visual senses is ``disruptive in predominantly sighted environments.'' 


\revision{Participants mentioned relying the least on their sense of smell for exploration.} Many recalled specific examples: such as detecting the smell of fries (Anya), laundry (\revision{Zendaya}), and garbage (\revision{Rahul}) to identify certain landmarks, helping them make an educated guess about the possible layout of the space. \revision{However, this information is not concrete enough for them to act upon, and is primarily used to supplement other senses.} \vspace{-0.25cm}\\\\
\begin{tabu} to \linewidth { X[l] X[28l] X[l]}
 & %
 \textit{``That's usually not my first priority. [The sense of smell] is more maybe just providing some supplemental information. I would not say that that's something that I rely on in any capacity.''} --\revision{Rahul}
 & \\
\end{tabu}\\



\vspace{2mm}
\vspace{-0.0cm}\begin{center}
\setlength{\fboxsep}{0.8em}
\fbox{\begin{minipage}{0.5\textwidth}
\textbf{Insight 3}: VIPs face difficulties in making \mbox{navigation} decisions based on spatial information collected via their non-visual senses.
\end{minipage}}
\end{center}
\vspace{1mm}

\subsubsection{Acquiring appropriate O\&M training and maintaining O\&M skills} 
\label{sec:Acquire-O&M-Training}
We discovered that most participants (n=7) expressed difficulties in exploring environments independently because of lack of confidence in their O\&M proficiency. Participants indicated that this was not because O\&M techniques are not sufficient to support independent exploration, but rather attributed this hesitation to their lack of access to appropriate O\&M training and their inability to maintain O\&M skills.

Upon further inquiry, we found three hindrances that prevent VIPs from acquiring adequate O\&M training resources: lack of funding, shortage of O\&M instructors, and difficulties in setting up O\&M sessions as an adult --- specifically those who developed vision impairments later in life. 
Most participants were aware of the power of O\&M skills and lamented on the lost opportunities to explore unfamiliar environments by themselves. 
As \revision{Charles} exclaimed: \vspace{-0.25cm}\\\\
\begin{tabu} to \linewidth { X[l] X[28l] X[l]}
 & \textit{``I feel like I would have benefited [with more O\&M training]. I don't want to feel like I'm over exaggerating, but at least 10 times more.''} --\revision{Charles} & \\
\end{tabu} \\%

Few participants (n=3) reported receiving appropriate O\&M training, yet still did not feel confident in exploring environments independently. Participants noted that this was because of their inability to regularly use their O\&M skills safely, and thus were concerned about atrophy of their skills.
\revision{Booker} elaborated how he denies assistance out of fear of his skills deteriorating if he relies too much on others.\vspace{-0.25cm}\\\\
\begin{tabu} to \linewidth { X[l] X[28l] X[l]}
 & \textit{"It is very easy to let your [O\&M] skills atrophy as a blind person if you rely overly on sighted people[.] It's easy to [use a] sighted guide and become complacent and [not pay attention to] what turns you have taken and where you are in relation to other things." } --\revision{Booker} & \\
\end{tabu}
\\\\However, he mentioned eventually relying on sighted guide assistance --- especially when exploring unfamiliar environments.

\vspace{2mm}
\begin{center}
\setlength{\fboxsep}{0.8em}
\fbox{\begin{minipage}{0.5\textwidth}
\textbf{Insight 4:} VIPs face difficulties in acquiring \mbox{appropriate} O\&M training and maintaining their O\&M skills, negatively impacting their confidence to  explore independently.
\end{minipage}}
\end{center}
\vspace{1mm}

\subsection{Challenges in Collaborative Exploration}
\label{sec:colab_exploration}
We discovered two major challenges that VIPs face when exploring unfamiliar environments by seeking assistance from other people. 
First, our VIP participants (n=4) expressed concerns about putting social pressure on others while seeking assistance from them. 
Second, many VIPs (n=5) described difficulties in supervising assistance from others, \revision{which is important to their sense of independence}. 
We elaborate on these challenges in the following sections.





\subsubsection{Social pressures in collaborative exploration}
Participants cited instances where they felt uncomfortable receiving help from others, due to social pressure that the person helping them may face. 
\revision{Booker} expressed frustrations about having to put her younger sister in a position where others feel that she is responsible for \revision{Booker}'s actions. He recalled going for shopping with his sister: \vspace{-0.25cm}\\\\
\begin{tabu} to \linewidth { X[l] X[28l] X[l]}
 & \textit{“I know that my sister gets looks from people all the time, because of foolish assumptions they make out of ignorance. [People think] that she is my caretaker [and that] anything that they perceive as her not doing her job, is wrong.”} --\revision{Booker} & \\
\end{tabu}%
\\\\
Many participants (n=4) indicated that they are most comfortable seeking assistance from O\&M instructors since they are trained professionals and somewhat immune to these social pressures. However, relying on O\&M instructors is not always possible, as we discussed earlier (Section~\ref{sec:Acquire-O&M-Training}).


During our interviews, most VIPs (n=10) recalled examples of grocery stores as an example setting for collaborative exploration. Participants described several different techniques for buying groceries. \revision{Daniel}, for instance, stated that he calls the store ahead of time to arrange for a personal shopper. Despite the advance notice for assistance, \revision{Daniel} describes sensing some discomfort when exploring the store with the store employee. To better understand this discomfort and to confirm \revision{Daniel}'s inkling, we followed up with a grocery store employee.

We found that the employees indeed feel the social anxiety, confirming \revision{Daniel}'s claim. 
The grocery store employee noted that despite his intention to help visually impaired customers, he is \textit{``... always concerned with invading somebody's personal space or [disregarding] their bodily autonomy.''} He also mentioned worrying about the company policies and feared getting in trouble if he makes a mistake. \vspace{-0.25cm}\\\\
\begin{tabu} to \linewidth { X[l] X[28l] X[l]}
 & \textit{``[My manager] said that if you need to help a customer [...] who's disabled, definitely try to if they give you consent [...] Just be careful in that case. But if they don't say anything, then you shouldn't do anything. Because you don't want to actually [cause] issues where they say, `oh, why are you touching me'?''} --Grocery store employee & \\
\end{tabu} \\\\
We discovered that social pressures affect both VIPs and others, negatively impacting the frequency and quality of assistance that VIPs receive when exploring environment collaboratively. 

\vspace{2mm}
\begin{center}
\setlength{\fboxsep}{0.8em}
\fbox{\begin{minipage}{0.5\textwidth}
\textbf{Insight 5}: Social pressures pose a major challenge to VIPs receiving help and to non-VIPs providing help when exploring environments collaboratively.
\end{minipage}}
\end{center}
\vspace{1mm}


\subsubsection{Supervising assistance for collaborative exploration}
\label{sec:supervising-collab-exploration}

Many participants (n=4) described wanting a loose form of collaboration, specifically verbal assistance, when exploring unfamiliar environments with the help of others. Verbal assistance comprises primarily of descriptions for shape and layout information, which enables them to \textit{``figure things out''} by themselves (\revision{Zendaya}). 
\red{By contrast, Williams et al.~\cite{williams_just_2014} found that VIPs prefer ``sighted guide'' assistance when walking with another person. \revision{This difference highlights that for VIPs, exploration is a self-driven activity and other people play a secondary role in this collaboration.}}
Our participants described challenges in communicating preference for verbal assistance. 




In addition, participants mentioned  challenges in comprehending verbal assistance. In situations where VIPs do get verbal assistance, they noted that the information they received was often not useful. We found that participants struggle to comprehend spatial information received from non-VIPs because of non-VIPs' predominantly visual descriptions.
\vspace{-0.25cm} \\\\
\begin{tabu} to \linewidth { X[l] X[28l] X[l]}
 & \textit{``[I need to ask for clarification] because a lot of people describe [shapes with letters] like `U' shaped or `H' shaped. ... I just prefer geometric shapes [for descriptions.]''} --\revision{Booker} & \\
\end{tabu} \vspace{-0.1cm}\\\\
\red{While several studies have noted that VIPs' perceptions of space is different from those of sighted people~\cite{williams_just_2014, thieme_i_2018}, we identified VIPs' preferred format for spatial information --- geometric shape descriptions --- to inform NAS design.}

Interestingly, we found that for this reason many participants (n=3) collaborate with other VIPs who have already been to a location. \revision{Zendaya} recounted calling her friend, who is also visually impaired, to ask for verbal assistance. Through her friend she obtained spatial information in a more useful format, along with additional cues for her non-visual senses. \vspace{-0.25cm}\\\\
\begin{tabu} to \linewidth { X[l] X[28l] X[l]}
 & \textit{“So for example, my friend will tell me this train [...] doesn't talk. And I'm really gonna have to pay attention and count the stops.”} --\revision{Zendaya} & \\
\end{tabu} \\\\
Another successful instance of such a collaboration is using remote sighted assistance such as AIRA~\cite{aira}. Although participants are content with verbal descriptions provided by AIRA professionals, many participants (n=5) noted that it is too expensive for them.



\vspace{2mm}
\begin{center}
\setlength{\fboxsep}{0.8em}
\fbox{\begin{minipage}{0.5\textwidth}
\textbf{Insight 6}: VIPs want verbal assistance in a \mbox{comprehensible format} when exploring \mbox{collaboratively} but find it challenging to \mbox{communicate this} preference to others.
\end{minipage}}
\end{center}
\vspace{2mm}

\section{Discussion}
\label{sec:DGs}

Based on our finding's key insights (Insights 1--6), we present design implications for NASs that can assist VIPs in exploring unfamiliar environments (Section~\ref{sec:design-implications}). We hope that these design implications can serve as a catalyst for future research in realizing such NASs. 
Additionally, in Section~\ref{sec:exploration-vs-turnByturnNav}, we elaborate on how best to balance turn-by-turn navigation with exploration-based navigation. Specifically, we discuss the interplay between these two forms of navigation and describe an opportunity for accessibility research to focus on \textit{how} users accomplish tasks and not simply \textit{whether} they can accomplish tasks.
\revision{
Lastly, we reflect upon our study procedure with an aim to inform the CSCW community on ethical considerations that may be adopted by future research (Section~\ref{sec:ethical_cons}).
}

\begin{table}[t]
\renewcommand{\arraystretch}{1.2}
\caption{Summary of our study findings in terms of Insights 1--6 and design implications for NASs that incorporate exploration assistance.}
\label{tab:findings}
\centering
\resizebox{\textwidth}{!}{%
\begin{tabular}{{p{0.145\textwidth}p{0.41\textwidth}p{0.41\textwidth}}}
\toprule
 \hspace{0.4em}\textbf{Theme} & \textbf{Insight} & \textbf{Design Implication}\\
\midrule


\multicolumn{1}{l}{\multirow{2}{*}{\begin{tabular}{{p{0.145\textwidth}p{0.41\textwidth}p{0.41\textwidth}}}
\vspace{-0.6cm}Information needs
\end{tabular}}}
& \textbf{I1:} VIPs need two types of spatial information: shape information and layout \mbox{information}, to get a high-level overview of a space.
& NASs should help VIPs gather shape and layout information in a manner that \mbox{facilitates} active engagement with the environment.
\\

\hline

\multicolumn{1}{l}{\multirow{2}{*}{\begin{tabular}{{p{0.145\textwidth}p{0.41\textwidth}p{0.41\textwidth}}}Factors \mbox{influencing} information needs\end{tabular}}}
& \textbf{I2:} VIPs’ preferred format, source, and amount of spatial information support vary between individuals depending upon their onset of vision impairment, their inherent sociability, and their O\&M proficiency and mobility aid preference, respectively.
& NASs should collect spatial information in a manner that allows VIPs to customize its display along the three dimensions of format, source, and amount of spatial information support.
\\
\hline


\multicolumn{1}{l}{\multirow{2}{*}{\begin{tabular}{{p{0.145\textwidth}p{0.41\textwidth}p{0.41\textwidth}}}Independent exploration\end{tabular}}}
& \textbf{I3:} VIPs face difficulties in making \mbox{navigation} decisions based on spatial \mbox{information} collected via their non-visual senses.
&  NASs should afford VIPs precise and \mbox{reliable} spatial information and should ensure that VIPs  make accurate navigation decisions based on this information. \\

\cline{2-3}

& \textbf{I4:} VIPs face difficulties in acquiring appropriate O\&M training and maintaining their O\&M skills, negatively impacting their confidence to explore independently.
& NASs can serve as an O\&M education \mbox{assistance} tool to give VIPs the \mbox{confidence} to explore unfamiliar environments independently.
\\


\hline

\multicolumn{1}{l}{\multirow{2}{*}{\begin{tabular}{{p{0.145\textwidth}p{0.41\textwidth}p{0.41\textwidth}}}Collaborative exploration\end{tabular}}}
& \textbf{I5:} Social pressures pose a major challenge to VIPs receiving help and to non-VIPs providing help when exploring environments collaboratively.
& 
NASs should normalize VIPs' exploration behaviors 
by introducing social norms for exploration assistance. 
\\

\cline{2-3}
 
& \textbf{I6:} VIPs want verbal assistance in a comprehensible format when exploring collaboratively but find it challenging to communicate this preference to others.
& NASs should scaffold and facilitate collaborative exploration by distributing \mbox{requests} for collaboration and \mbox{performing} translations when needed.
\\

\bottomrule
\end{tabular}
}
\end{table}

\subsection{Design Implications for NASs}
\label{sec:design-implications}
\Cref{tab:findings} summarizes key insights from our findings and design implications for developing future NASs.
The interviews helped us identify key insights about VIPs' information needs, factors that affect these information needs, and challenges that VIPs face that impede their ability to explore unfamiliar environments. In the following sections, we describe how NASs can be designed to support these information needs and to alleviate these challenges. \\

\subsubsection{Facilitating active engagement and supporting individual variations}

Our findings reveal that VIPs need two types of spatial information --- shape information and layout information --- in order to obtain a high-level overview of a space (Insight 1). Participants described wanting to acquire this spatial information before anything else (Section~\ref{sec:informationNeeds}), and recent research~\cite{clemenson_environmental_2017,MSsoundscape,Nair2021a} has shown that VIPs should be able to gather this information via interactions that foster active engagement.
Our findings also reveal the importance of allowing VIPs to customize the format, source, and amount of spatial information support they receive to suit their background (Insight 2).

\begin{figure}[t]
    \centering
    \includegraphics[width=0.7\linewidth]{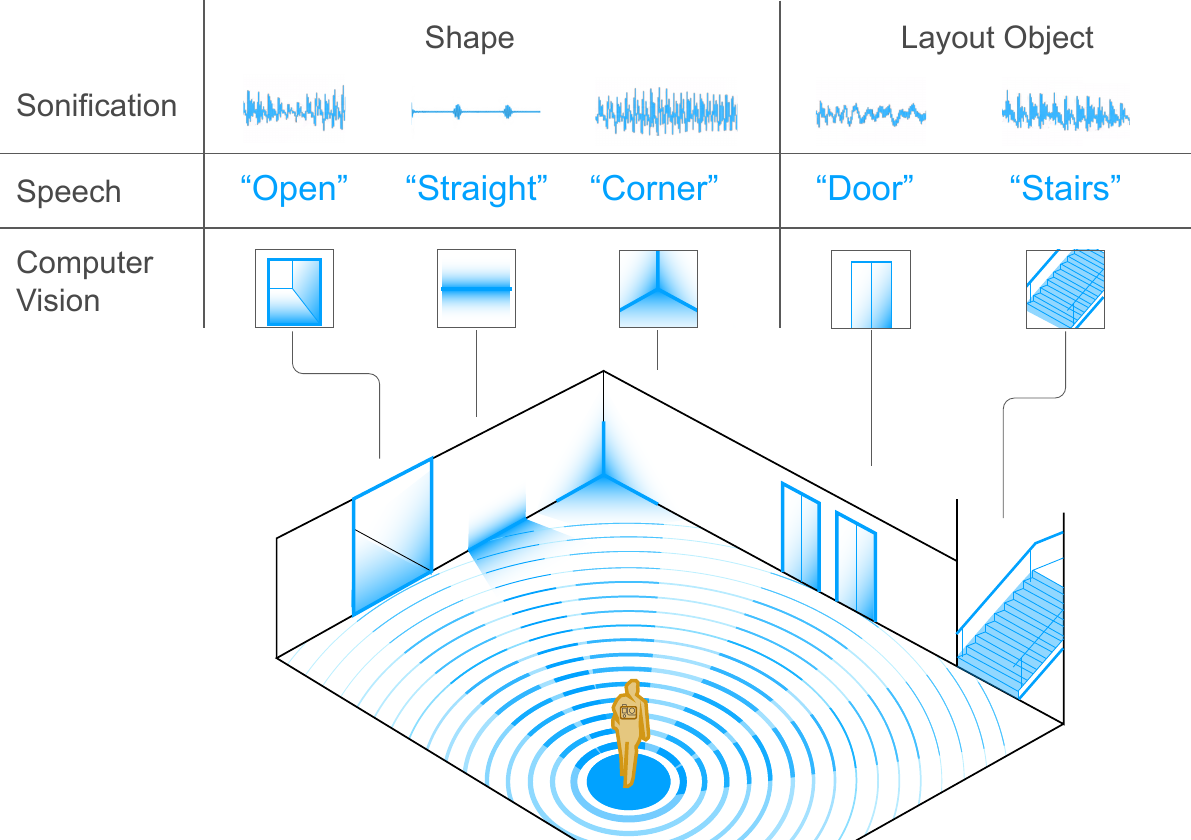}
    \caption{Conceptual illustration of an NAS facilitating spatial information in \revision{an indoor environment}. The NAS consists of a wearable sensor that detects the environment's \textit{shape} and \textit{layout} using computer vision. Users can actively explore by ``probing'' the environment on demand and can choose a preferred format for the information: either sonified waveforms or speech. We found that early-blind and late-blind users respectively prefer these two formats. \revision{Similarly, we can extend this concept to facilitate spatial information in outdoor environments as well. We can imagine such a similar system being used in outdoor environments that detects sidewalks, crosswalks, pavements, and the shape of walkable area within the space.}}
    \label{fig:spatialInfo}
\end{figure}

Collectively, these insights can influence NAS design in exciting new ways.
In Figure ~\ref{fig:spatialInfo}, we show a conceptual illustration of how a future NAS might help users gather spatial information \revision{in an indoor environment} while facilitating both active engagement and customization.
\revision{
We note that the illustration we propose is one of the many possibilities in the design space of building such NASs. Through this example, we aim to kick-start discussions around socio-technical challenges and practical considerations that the research community would need to address before realizing NASs that support exploration (more on this in Section~\ref{sec:challenges}).
}

\revision{To support exploration, we conceptualize} a wearable sensor such as LiDAR or a depth camera that scans for shape information by recognizing different elements of the environment including corners, straight edges, and curved edges. The sensor also scans for layout information by detecting objects such as doors, stairs, and elevators.
The NAS conveys this information to the user via speech or sonified waveforms depending on the user's preference --- recall from Section~\ref{sec:spatial-info-format} that early-blind VIPs prefer amplifying their ability to hear while late-blind VIPs prefer spoken descriptions. 
Users receive this feedback on-demand, perhaps by ``looking around'' via a NavStick-like interface~\cite{Nair2021a}, since continuously ``pushed'' feedback can impair VIPs' ability to hear safety-critical sounds~\cite{banovic_uncovering_2013, williams_just_2014, easley_lets_2016}. \revision{We can imagine a similar system being used in outdoor environments, detecting sidewalks, crosswalks, pavements, and the shape of walkable area within the space.} 

\revision{
Beyond wearable sensor-based NASs, navigation tools such as tactile maps have potential to help VIPs gather shape and layout information from an allocentric reference~\cite{espinosa_using_1998, blades_map_1999, trinh_feeling_2020, abd_hamid_facilitating_2013, jacobson_navigating_1998, taylor_customizable_2016, miesenberger_pre-journey_2014, nadel_hippocampus_1980}. However, in order to render graphics on tactile maps, information about the floor plan (indoors) or street view maps (outdoors) is required, which may not always be available for unfamiliar environments. 
Combining wearable sensors with tactile maps provides opportunity to support VIPs' exploration by sensing VIPs' immediate environment and rendering it via tactile maps in real-time. Recent advancements in commodity tactile displays such as the Dot Pad~\cite{dotpad}, would further allow researchers to develop tools that may support different use cases using the same device.
Although sensing systems that use braille displays are not uncommon~\cite{wang_enabling_2017}, they mostly provide route-based knowledge such as obstacles and landmarks. 
Re-purposing existing tools and systems to support exploration is an open area for research with its own challenges. Dey et al.'s work~\cite{dey_getting_2018, dey_understanding_2019} on designing map-based interfaces that support exploration for sighted people provides a good starting point for future research.
}

\revision{
\subsubsection{Challenges and considerations in supporting exploration via NASs}
\label{sec:challenges}
While our work introduces several ideas for how NASs can support VIPs' exploration, future researchers must still solve technical challenges, be cognizant of practical considerations, and address social considerations before such systems can make a real impact in VIPs' lives. We discuss these challenges in context of the wearable sensor-based NAS illustrated in Figure~\ref{fig:spatialInfo}.
}

\revision{
Research in robot navigation and computer vision has made significant progress in perceiving the environment using navigation stacks~\cite{guerreiro_cabot_2019} and deep learning models~\cite{lin_deep_2019, bauer_enhancing_2020}. 
However, systems for supporting exploration such as the concept expressed in Figure~\ref{fig:spatialInfo} requires further research and development. 
Technical challenges may be caused by several things: (1) occluding objects or other people within the space: making it hard to perceive the actual shape of the environment, (2) finite range of sensors: limiting applicability of such systems to large open spaces such as shopping malls, and (3) camera instability of on-body sensors: causing blur and severely affecting the accuracy of vision algorithms as the user walks around.
%
%
%
%
}


\revision{
In addition to the technical challenges, NAS designers would need to address several practical considerations. To begin with, NASs should not be cost-prohibitive for VIPs, who have been found to earn 30--40\% less than sighted people~\cite{kirchner1999looking, American2017Interpreting, frick_economic_2007}. Low-cost sensing methods such as pure camera-based systems, however, are typically much less accurate than systems such as LiDARs.
It will be important for researchers to develop calibration and error-recovery mechanisms to counter the limited capabilities of low-cost equipment. 
Yoon et al.'s smartphone-based NAS~\cite{yoon_leveraging_2019}, for example, require users to initialize it by orienting the phone horizontally to face the ground, always at the same location in the beginning of a session. 
}


\revision{
We also found that VIPs do not want to ``stand out'' while navigating unfamiliar environments, wanting a more discreet form-factor for the wearable sensor such as that depicted in Figure~\ref{fig:spatialInfo}. Interestingly, this finding contradicts with Lee et al.'s recommendation that NASs should be visible to other people within the environment, in order to assuage their privacy concerns. Future approaches should further explore this contradiction and begin to find ways for managing the trade-off.}

\revision{
One solution may be to create awareness within the sighted community about common assistive technology devices. K-12 schools could, perhaps, include lessons on accessibility by discussing VIPs' general practices and use of assistive technology to instill acceptance and awareness. Another solution could be introducing new security mechanisms within NASs to further ease this conflict. For instance,
NASs such as that depicted in Figure~\ref{fig:spatialInfo} often send data over cloud for complex computations of computer vision algorithms. To prevent leakage of any private information, future NASs could perform the initial feature extraction on-device, before any data is sent over the network for processing. This would reduce the risk of data leakage over the network.}


\subsubsection{Precision and reliability afford in-situ exploration}

From our interviews, we learned that VIPs hesitate in acting upon information collected through their non-visual senses such as hearing, touch, and smell~(Insight 3). For example, both Charles and James expressed concerns about making navigation decisions based on the rough estimate of an \revision{area}'s shape that they gathered via echolocation. Hence, NASs should provide precise and reliable spatial information about environments to VIPs. Just as turn-by-turn navigation efforts have spurred research on accurately locating users and providing them with precise turn-by-turn instructions~\cite{murata_smartphone-based_2018, ahmetovic_achieving_2017, ahmetovic_navcog_2016}, 
the cause of facilitating exploration assistance will require research on accurately sensing environments' shapes and layouts and providing this information to users in a precise manner.

\begin{figure}
    \centering
    \includegraphics[width=0.5\textwidth]{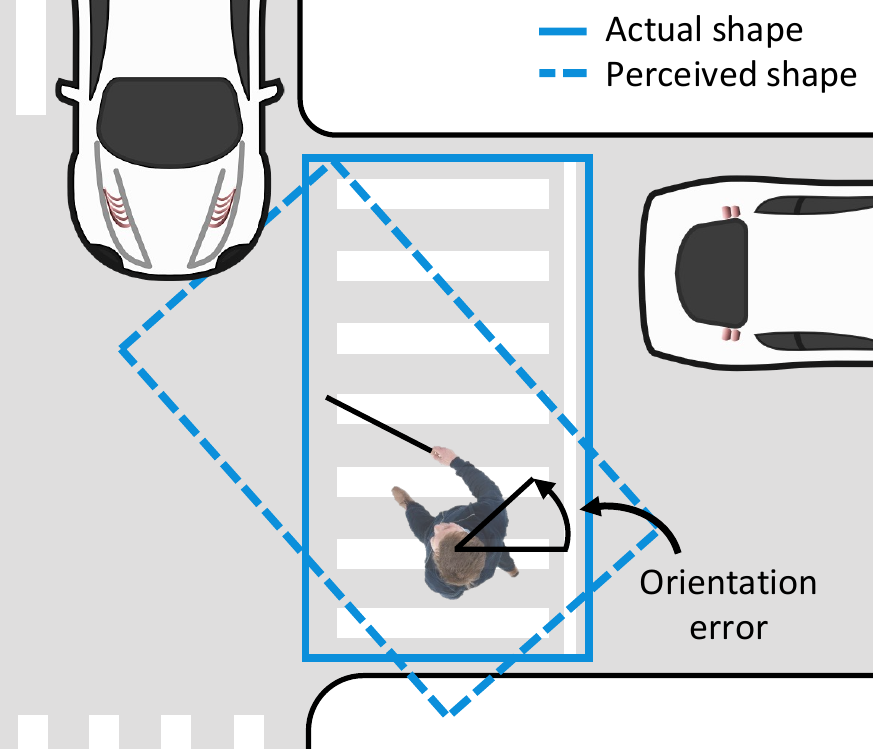}
    \caption{Illustration of a rotationally inconsistent cognitive map \revision{in an outdoor environment. Here, a VIP's imprecise sense of orientation leads to them incorrectly understanding the how a crosswalk is situated. This may lead them in the path of a moving car unless they course-correct using the tactile information gathered via their white cane or the sound of the cars and other pedestrians.} Future NASs may need to identify and resolve these types of user errors during in-situ exploration.}
   \label{fig:rotational-precision}
\end{figure}

In addition, future NASs may have to perform post-hoc interventions to compensate for impreciseness that may arise in VIPs' perception of their surroundings. 
For instance, a slight imprecision in VIPs' sense of orientation could lead to rotationally inconsistent cognitive maps of the space, causing VIPs to make incorrect navigation decisions. As an example, in Figure~\ref{fig:rotational-precision} we illustrate how a minor error in VIP's perceived orientation could lead them to misjudge \revision{the walkable area within an outdoor environment, a crosswalk in this case, and potentially end up in a dangerous situation.} 
Many researchers have recently focused on mitigating
user errors such as veering~\cite{pan_walking_2013, prattichizzo_anti-veering_2018, mandanici_following_2018} and over-turns~\cite{ahmetovic_turn_2018, ahmetovic_deep_2020} because of their affect on how well turn-by-turn navigation assistance systems perform.
Analogously, we can imagine further research into identifying and resolving user errors that arise during in-situ exploration assistance.


\subsubsection{O\&M education assistance as a means of empowering VIPs to explore}
\label{sec:onm-tools}
From our interviews, we learned that VIPs lack confidence in their O\&M skills --- due to either their lack of appropriate O\&M training or lack of opportunities to maintain their skills --- which leads many VIPs to become wary of exploring new places even if they may want to. Participants agreed that having appropriate O\&M training and opportunities to practice their O\&M skills regularly would enable them to independently explore. In short, O\&M skills are a powerful enabler for exploration.

As a result, immense opportunities arise for NASs to be designed in a way that 
promotes O\&M education as users navigate, empowering VIPs to become confident in exploring environments. 
In other words, future NASs can include built-in O\&M education facilities to give VIPs access to O\&M instruction and support without requiring human instructors, which we found in Section~\ref{sec:Acquire-O&M-Training}) to be in very short supply. Very recently, Harriman et al.~\cite{harriman_clew3d_2021} introduced Clew3D, a smartphone application that uses LIDAR to automatically generate O\&M instructions for a route, identifying O\&M landmarks such as open areas, walls that can be shorelined, and doors as part of the instructions.
This work takes an exciting step toward incorporating O\&M training into NASs, but it is turn-by-turn-focused and does not explore facilities for O\&M teaching or practice. 

Future NASs can further incorporate O\&M education by offering O\&M training modules and posing O\&M training ``challenges'' or ``problem sets'' over the course of the day if users want them, doing so in a similar manner to education and workout apps. 
The NASs can craft these challenges for the user based on the systems' own understanding of the environments' shape and layout. They should be careful, however, not to push users into taking up tasks they do not wish to practice or perform --- users should be in control of when and how they would like to learn. 
O\&M instructors themselves can also be part of the loop --- one can imagine an NAS that keeps detailed system logs of a user's O\&M performance while navigating and then forwarding the logs to an O\&M trainer for review.
The O\&M trainer might use these logs to identify O\&M techniques that the user should practice more.


\subsubsection{Mitigating social pressures by standardizing assistance for exploration}

VIPs' navigation success is known to be affected by both internal and external factors~\cite{williams_pray_2013}. 
Our findings (Insight 5) reveal that social pressures often arise when other people are present and when others may perceive both the VIP and any people helping the VIP negatively. These social pressures can dramatically hurt VIPs' ability to receive help and collaboratively explore. 
Moreover, our follow-up interviews with relevant non-VIPs revealed that even people helping VIPs feel social anxiety about how VIPs and onlookers perceive their ability to help. VIPs are aware of this effect on people helping them and avoid putting them in that position.

Part of this problem, as we learned from Harper's example of people looking at VIPs strangely if they were to get up at a restaurant and start feeling all of the tables (Section~\ref{sec:precise-spatial-info}), is that it is currently much less socially acceptable for VIPs to explore and receive exploration assistance from others than it is for VIPs to be guided and receive ``sighted guide'' help from others. Future work can mitigate this problem in two ways. First, NASs with exploration capabilities can enable VIPs to explore environments' shapes and layouts in a much more private way, perhaps through use of a smartphone app rather than having to physically touch objects.

Second, future efforts and advocacy can work to normalize exploration behaviors and establish social norms for exploration that are analogous to the sighted guide technique.
\revision{To create awareness about these new standards for exploration, we should leverage social media and internet like many disability activists.} 
Future efforts can also promote awareness and empathy by building upon prior work on emulation software~\cite{flatla_so_2012, hailpern_aces_2011} that can simulate how people with disabilities ``see'' or ``feel.''

\subsubsection{Facilitating effective collaboration to enable further exploration}
\label{sec:exploring-with-ppl}

Our interviews revealed that collaboration is a powerful enabler of exploration for VIPs and that VIPs' ability to collaborate effectively with other people (including other VIPs) can help them explore unfamiliar environments further. Participants noted two issues in collaborating effectively with others: issues with communicating their preference for verbal assistance and issues with comprehending others' verbal assistance (Insight 6). 
As a result, two opportunities emerge for exploring how NASs can bridge communication between VIPs and others to unlock the potential that collaboration can yield. 

First, future NASs might help scaffold and facilitate the process of
requesting assistance from others, allowing VIPs to specify what information they need as well as what format they need this information.
Researchers can explore how NASs might issue structured requests for exploration assistance to nearby people when VIPs ask for it.
Second, with respect to comprehending verbal assistance, future NASs might employ natural language processing (NLP) to translate verbal assistance from others to VIPs' preferred format. In Section~\ref{sec:supervising-collab-exploration}, we found a mismatch between the languages that VIPs and others use during collaboration. 


Of course, important ethical concerns arise from people responding to requests from an NAS rather than a person right next to them. Perhaps, NASs should play a role similar to remote sighted assistance systems such as AIRA~\cite{aira}, which our participants complained as being cost-prohibitive to them (Section~\ref{sec:supervising-collab-exploration}). Instead of experts answering VIPs' questions, future research may explore how NASs could do the same to save costs and remove barriers to access. Researchers could collect data from conversations between AIRA agents and VIPs to feed as training data into the NLP algorithm and could investigate the real-life performance enabled by such training. More research can explore how AIRA agents and other VIPs can be kept in the loop to moderate translations in real time.

\subsection{Balancing Exploration Assistance with Turn-by-Turn Guidance}
\label{sec:exploration-vs-turnByturnNav}


From our interviews, we discovered that VIPs \textit{``want to figure things out''} (Nora) themselves and desire the ability to explore unfamiliar environments in-situ. 
While all participants appreciated this ability, some (n=3) did not express a strong preference for having it. These participants indicated their satisfaction with being guided via turn-by-turn instructions, explaining that it was all they needed for their daily-life activities, reinforcing the importance of prior work that builds from this perspective~\cite{ahmetovic_navcog_2016, fallah_user_2012, bai_virtual-blind-road_2018,murata_smartphone-based_2018, ran_drishti_2004, riehle_indoor_2008, guerreiro_cabot_2019, kayukawa_bbeep_2019, kayukawa_guiding_2020, avila_soto_dronenavigator_2017, peng_indoor_2017, saegusa_development_2011, scheggi_cooperative_2014}. 

In fact, even participants who wish to actively explore environments shared specific instances in which they did not mind being guided to a destination. Lucas and William, for example, preferred to be guided at airports when time and convenience were of essence to them. We also observe a clear overlap between VIPs' information needs for exploration and for turn-by-turn navigation. As discussed in Section~\ref{sec:informationNeeds}, VIPs described wanting spatial information --- information needed for exploration --- \textit{in addition} to route-based information, which is needed for turn-by-turn navigation.

Since VIPs expressed a desire to explore environments in some situations and be guided in a turn-by-turn fashion in other situations, we foresee future NASs facilitating both forms of navigation.
However, several questions still remain unanswered. How exactly should NASs combine the two forms of navigation? When do VIPs want to explore and when do they want turn-by-turn instructions? Which agent --- the user or the NAS --- should initiate the switch between the two forms of navigation, and what factors might NASs use to determine the level of initiative~\cite{Horvitz1999} they should take in a given situation? While extensive research is needed to comprehensively answer these questions, our findings on VIPs' exploration preferences and the prior body of research in turn-by-turn navigation can serve as starting points for future investigations.

Ultimately, the significance of exploration and users' sense of agency highlights
the importance of framing navigation assistance and other assistive technologies as not just tools but as \textit{experiences} for users. 
Assistive technology research for a range of populations can benefit from 
focusing on users' experience while using the system in addition to what functionality the system can help users accomplish. This means that evaluations should weigh both functionality (i.e., whether users can complete the task) and experience (i.e., how users' actual roles compare to the role they want to play).

\revision{
\subsection{Ethical Reflections of Our Study Procedure}
\label{sec:ethical_cons}
In line with recent trends in accessibility research~\cite{lee_designing_2021, chase_pantoguide_2020, yoon_leveraging_2019, cullen_co-designing_2019, siu_shapecad_2019, bennett_interdependence_2018}, we found it crucial that VIPs are represented in research not only as study participants but also as co-interpreters of the data. In fact, for certain parts of our data, we found it helpful to consult with the VIP interns (involved in the capacity of co-authors on another project within our lab) for quick clarifications on our understanding of the study data. Through this process of including VIPs in the interpretation process, we discovered that the insights and clarifications from the VIP interns stemmed from not just their own lived experiences~\cite{polanyi_tacit_1997}, but also from experiences of their VIP friends. 
For instance, one of our low vision VIP intern often cited experiences of their totally blind friend during our conversations. Future research approaches should consider including representative users for interpreting study data, especially for research in accessibility. \mnrevision{Involving representative users in different roles also calls for further research to better understand and establish the best practices for ensuring appropriate compensation and due credit for the representative users' contributions~\cite{lundgard_sociotechnical_2019}.}} 

\revision{
%
Additionally, we note that using CIT~\cite{flanagan_critical_1954} during the interviews leads to discussion around actual examples of participants' past experiences, which sometimes may include private identifiable information such as locations of places participants visit, names of their friends and family members. Although we separated participant's name and contact information from their study data, some private identifiable information may have, unintentionally, been stored as part of the interview transcripts. To maintain confidentiality, we redacted all such private information while reviewing the transcripts. 
Future approaches using CIT should consider reviewing the study data to ensure confidentiality. 
}

\section{Limitations}

\label{sec:limitations}
While our work provides key insights on how NAS designers might rethink the design of NASs to support exploration in addition to turn-by-turn navigation, we acknowledge that our study has several limitations. 
The study sample included only 12 visually impaired participants, all of whom live in the United States and whose navigation preferences and experiences may not be representative of the visually impaired community at large. 
Since recruiting participants with disabilities is a perrenial challenge faced by the accessibility research community, there is an increased understanding among the community to recruit smaller samples from local venues through community-based organizations~\cite{sears_representing_2012} and snowball sampling~
\cite{goodman_snowball_1961}, which is the approach that we took. 
Still, we must acknowledge this limitation nonetheless. In the future, our study can be extended to elicit responses from a large sample of VIPs from different countries to garner insights from more diverse socio-cultural experiences.

Another limitation arises from our use of CIT~\cite{flanagan_critical_1954} in our interviews with participants. Namely, we relied on participants' memory of their past experiences and behaviours, which may not always be comprehensive or accurate. 
Although our approach is the first to explore how NASs can best address VIPs' challenges regarding exploration in navigation, a longitudinal study such as a diary study may yield insights that we may not have been able to encounter. \revision{More specifically, future research could shadow VIPs to understand the context of their experiences in more detail and reveal more subtle nuances that may have been overlooked by using CIT.}
Last, we do not yet know how our study results might generalize to other types of impairments. Thus, further research is needed to confirm and adapt some of our results to offer a more comprehensive picture of how assistive technologies might facilitate exploration for a variety of populations.

\section{Conclusion}
In this paper, we conducted semi-structured interviews with 12 VIPs and others who affect VIPs' navigation behaviors to investigate VIPs' information needs and challenges with respect to exploring unfamiliar environments. The broader goal was to establish a holistic understanding of how NASs can be designed to support exploration in addition to turn-by-turn navigation.
We discovered that VIPs need two types of spatial information: \textit{shape} and \textit{layout} information, to get a high-level overview of spaces and to make navigation decisions \textit{in situ}. 
VIPs' preference for the format, source, and amount of spatial information varies between individuals depending upon their onset of vision impairment, their inherent sociability, and their O\&M proficiency and mobility aid preference, respectively. 
Our findings also described specific challenges that VIPs face when exploring unfamiliar environments, both independently and in collaboration with other people. Among these are building confidence in orientation and mobility skills and collaborating effectively with others who might enable further exploration.

Throughout our findings, we highlight key insights that lead to important design implications for NASs supporting exploration.
We discuss several avenues for future research including how NASs might facilitate spatial information in a manner promoting active engagement and how NASs can serve as educational tools to help VIPs build confidence in their O\&M skills, thereby supporting VIPs' exploration in navigation. 
We hope that our work inspires the research community to adopt a more agency-oriented view of assistive technology design --- one that considers users' sense of self-fulfillment using assistive technologies 
to be as important as completing the task at hand. \revision{
Finally, we share ethical reflections of our study procedure to inform the future research approaches within the CSCW community.
}

\begin{acks}
We thank Michael Malcolm and Sebastián Mercado (referred to as VIP interns) for quick clarifications on
our understanding of the study data, Yilun Sun for the Fig.~\ref{fig:teaser} illustration, the reviewers for the feedback, and all our study participants.
\end{acks}

\bibliographystyle{ACM-Reference-Format}
\bibliography{ref.bib}


\end{document}